\pdfoutput=1

\documentclass[12pt,a4paper]{article}

\usepackage{graphicx}  
\usepackage{xspace} 
\usepackage{color}
\usepackage{colortbl}
\usepackage{amsmath} 
\usepackage{ifthen} 
\usepackage{subfigure}
\usepackage{overpic}
\newboolean{pdflatex}
\setboolean{pdflatex}{true} 
\newboolean{articletitles}
\setboolean{articletitles}{true} 

\newboolean{uprightparticles}
\setboolean{uprightparticles}{false} 

\usepackage{amssymb}
\usepackage{amsfonts}
\usepackage{upgreek} 

\usepackage{hyperref}    
\usepackage[all]{hypcap} 

\usepackage{booktabs}

\usepackage{xcolor} 
\definecolor{halfgray}{gray}{0.55} 
\definecolor{webgreen}{rgb}{0,.5,0}
\definecolor{webbrown}{rgb}{.6,0,0}
\definecolor{Maroon}{cmyk}{0, 0.87, 0.68, 0.32}
\definecolor{RoyalBlue}{cmyk}{1, 0.50, 0, 0}

\hypersetup{%
    colorlinks=true, linktocpage=true, pdfstartpage=3, pdfstartview=FitV,%
    breaklinks=true, pdfpagemode=UseNone, pageanchor=true, pdfpagemode=UseOutlines,%
    plainpages=false, bookmarksnumbered, bookmarksopen=true, bookmarksopenlevel=1,%
    urlcolor=webbrown, linkcolor=RoyalBlue, citecolor=webgreen, 
    pdftitle={Search for $D^0 \to mu^+ mu^-$ decays at LHCb},%
    pdfauthor={},%
    pdfsubject={},%
    pdfkeywords={},%
    pdfcreator={pdfLaTeX},%
  }

\def\lhcb {\mbox{LHCb}\xspace}
\def\ux85 {\mbox{UX85}\xspace}

\ifthenelse{\boolean{uprightparticles}}%
{

 \def\Ppsi        {\ensuremath{\uppsi}\xspace}

 \def\PDelta      {\ensuremath{\Delta}\xspace}                 
 \def\PXi      {\ensuremath{\Xi}\xspace}                 
 \def\PLambda      {\ensuremath{\Lambda}\xspace}                 
 \def\PSigma      {\ensuremath{\Sigma}\xspace}                 
 \def\POmega      {\ensuremath{\Omega}\xspace}                 
 \def\PUpsilon      {\ensuremath{\Upsilon}\xspace}                 
 
 \def\PB      {\ensuremath{\mathrm{B}}\xspace}                 
                  
 \def\PD      {\ensuremath{\mathrm{D}}\xspace}

 \def\PJ      {\ensuremath{\mathrm{J}}\xspace}                 
 \def\PK      {\ensuremath{\mathrm{K}}\xspace}

 \def\Pb      {\ensuremath{\mathrm{b}}\xspace}                 
 \def\Pc      {\ensuremath{\mathrm{c}}\xspace}

 \def\Pi      {\ensuremath{\mathrm{i}}\xspace}

}
{

 \def\Ppsi        {\ensuremath{\psi}\xspace}                 
                  
 \mathchardef\PDelta="7101
 \mathchardef\PXi="7104
 \mathchardef\PLambda="7103
 \mathchardef\PSigma="7106
 \mathchardef\POmega="710A
 \mathchardef\PUpsilon="7107
                  
 \def\PB      {\ensuremath{B}\xspace}                 
                  
 \def\PD      {\ensuremath{D}\xspace}

 \def\PJ      {\ensuremath{J}\xspace}                 
 \def\PK      {\ensuremath{K}\xspace}

 \def\Pb      {\ensuremath{b}\xspace}                 
 \def\Pc      {\ensuremath{c}\xspace}

 \def\Pi      {\ensuremath{i}\xspace}

}

\def\cquark    {\ensuremath{\Pc}\xspace}

\def\bquark    {\ensuremath{\Pb}\xspace}

\def\kaon  {\ensuremath{\PK}\xspace}
  \def\Kbar  {\kern 0.2em\overline{\kern -0.2em \PK}{}\xspace}

\def\Kz    {\ensuremath{\kaon^0}\xspace}
\def\Kzb   {\ensuremath{\Kbar^0}\xspace}
\def\KzKzb {\ensuremath{\Kz \kern -0.16em \Kzb}\xspace}
\def\Kp    {\ensuremath{\kaon^+}\xspace}
\def\Km    {\ensuremath{\kaon^-}\xspace}

\def\KpKm  {\ensuremath{\Kp \kern -0.16em \Km}\xspace}

  \def\Dbar    {\kern 0.2em\overline{\kern -0.2em \PD}{}\xspace}
\def\D       {\ensuremath{\PD}\xspace}

\def\Dz      {\ensuremath{\D^0}\xspace}
\def\Dzb     {\ensuremath{\Dbar^0}\xspace}
\def\DzDzb   {\ensuremath{\Dz {\kern -0.16em \Dzb}}\xspace}
\def\Dp      {\ensuremath{\D^+}\xspace}
\def\Dm      {\ensuremath{\D^-}\xspace}

\def\DpDm    {\ensuremath{\Dp {\kern -0.16em \Dm}}\xspace}

  \def\Bbar    {\kern 0.18em\overline{\kern -0.18em \PB}{}\xspace}

\def\jpsi     {\ensuremath{{\PJ\mskip -3mu/\mskip -2mu\Ppsi\mskip 2mu}}\xspace}

  \def\Y#1S{\ensuremath{\PUpsilon{(#1S)}}\xspace}

\def\Lbar {\ensuremath{\kern 0.1em\overline{\kern -0.1em\PLambda}}\xspace}

\def\to                 {\ensuremath{\rightarrow}\xspace}

\def\AT#1     {\ensuremath{A_{\mathrm{T}}^{#1}}\xspace}           

\def\C#1      {\ensuremath{\mathcal{C}_{#1}}\xspace}                       
\def\Cp#1     {\ensuremath{\mathcal{C}_{#1}^{'}}\xspace}                    
\def\Ceff#1   {\ensuremath{\mathcal{C}_{#1}^{\mathrm{(eff)}}}\xspace}        
\def\Cpeff#1  {\ensuremath{\mathcal{C}_{#1}^{'\mathrm{(eff)}}}\xspace}       
\def\Ope#1    {\ensuremath{\mathcal{O}_{#1}}\xspace}                       
\def\Opep#1   {\ensuremath{\mathcal{O}_{#1}^{'}}\xspace}                    

\def\dkpi       {\decay{\PD}{\PK\Ppi}}

\newcommand{\tev}{\ensuremath{\mathrm{\,Te\kern -0.1em V}}\xspace}
\newcommand{\gev}{\ensuremath{\mathrm{\,Ge\kern -0.1em V}}\xspace}
\newcommand{\mev}{\ensuremath{\mathrm{\,Me\kern -0.1em V}}\xspace}
\newcommand{\kev}{\ensuremath{\mathrm{\,ke\kern -0.1em V}}\xspace}
\newcommand{\ev}{\ensuremath{\mathrm{\,e\kern -0.1em V}}\xspace}
\newcommand{\gevc}{\ensuremath{{\mathrm{\,Ge\kern -0.1em V\!/}c}}\xspace}
\newcommand{\mevc}{\ensuremath{{\mathrm{\,Me\kern -0.1em V\!/}c}}\xspace}
\newcommand{\gevcc}{\ensuremath{{\mathrm{\,Ge\kern -0.1em V\!/}c^2}}\xspace}
\newcommand{\gevgevcccc}{\ensuremath{{\mathrm{\,Ge\kern -0.1em V^2\!/}c^4}}\xspace}
\newcommand{\mevcc}{\ensuremath{{\mathrm{\,Me\kern -0.1em V\!/}c^2}}\xspace}

\def\mum  {\ensuremath{\,\upmu\rm m}\xspace}

\def\invpb {\ensuremath{\mbox{\,pb}^{-1}}\xspace}

\def\invfb   {\ensuremath{\mbox{\,fb}^{-1}}\xspace}

\def\gsim{{~\raise.15em\hbox{$>$}\kern-.85em
          \lower.35em\hbox{$\sim$}~}\xspace}
\def\lsim{{~\raise.15em\hbox{$<$}\kern-.85em
          \lower.35em\hbox{$\sim$}~}\xspace}

\def\pt         {\mbox{$p_{\rm T}$}\xspace}

\def\evtgen     {\mbox{\textsc{EvtGen}}\xspace}
\def\pythia     {\mbox{\textsc{Pythia}}\xspace}

\def\geant      {\mbox{\textsc{Geant4}}\xspace}

\def\photos     {\mbox{\textsc{Photos}}\xspace}

\def\tell1  {TELL1\xspace}
\def\ukl1   {UKL1\xspace}

\def\dmumu{\mbox{\ensuremath{D^0 \to \mu^+ \mu^-}}\xspace}
\def\brmumu{\ensuremath{\mathcal{B}(D^0 \to \mu^{+} \mu^{-})}\xspace}
\def\dstdmumu{\mbox{\ensuremath{D^{\ast +}\to D^0 (\mu^{+} \mu^{-}) \pi^+}}\xspace}
\def\dstdpipi{\mbox{\ensuremath{D^{\ast +}\to D^0 (\pi^{+}\pi^{-}) \pi^+}}\xspace}
\def\peddstdmumu{{\ensuremath{\mu^{+} \mu^{-}}}\xspace}
\def\peddstdpipi{{\ensuremath{\pi^{+} \pi^{-}}}\xspace}
\def\peddstdpipimumu{{\ensuremath{\pi^{+} \pi^{-}\to \mu^{+} \mu^{-}}}\xspace}

\def\dpipi{\mbox{\ensuremath{D^0 \to \pi^{+} \pi^{-}}}\xspace}

\def\dkpi{\ensuremath{D^0 \to K^{-} \pi^{+}}\xspace}

\def\dstdkpi{\mbox{\ensuremath{D^{\ast +}\to D^0 (K^{-} \pi^{+}) \pi^+}}\xspace}

\def\dstdkmunu{\mbox{\ensuremath{D^{\ast +}\to D^0 (K^{-} \mu^{+} \nu_\mu)\pi^+}}\xspace}
\def\dstdpimunu{\mbox{\ensuremath{D^{\ast +}\to D^0 (\pi^{-} \mu^{+} \nu_\mu)\pi^+}}\xspace}

\def\dstdpi{\mbox{\ensuremath{D^{\ast +}\to D^0 \pi^+}}\xspace}

\def\Jmumu{\mbox{\ensuremath{J/\psi \to \mu^{+} \mu^{-}}}\xspace}
\def\deltamP{\mbox{\ensuremath{\Delta m_{\pi^{+}\pi^{-} }   }}\xspace}
\def\deltamM{\mbox{\ensuremath{\Delta m_{\mu^{+}\mu^{-}} }}\xspace}
\def\deltamK{\mbox{\ensuremath{\Delta m_{K^{-}\pi^{+}}}}\xspace}
\def\deltamhex{\mbox{\ensuremath{m_{h^{+}h^{(\prime)-}\pi^{+}}-m_{h^{+}h^{(\prime)-}} }}\xspace}
\def\deltamh{\mbox{\ensuremath{\Delta m_{h^{+}h^{(\prime)-}}}}\xspace}
\def\mP{\ensuremath{m_{\pi^{+}\pi^{-}} }\xspace}
\def\mM{\ensuremath{m_{\mu^{+}\mu^{-}} }\xspace}
\def\mK{\ensuremath{m_{K^{-}\pi^{+}} }\xspace}

\newcommand{\be}{\begin{eqnarray}}
\newcommand{\ee}{\end{eqnarray}}
\newcommand{\til}{~}

\usepackage{mciteplus}

\begin{document}

\renewcommand{\thefootnote}{\fnsymbol{footnote}}
\setcounter{footnote}{1}


\begin{titlepage}
\pagenumbering{roman}

\vspace*{-1.5cm}
\centerline{\large EUROPEAN ORGANIZATION FOR NUCLEAR RESEARCH (CERN)}
\vspace*{1.5cm}
\hspace*{-0.5cm}
\begin{tabular*}{\linewidth}{lc@{\extracolsep{\fill}}r}
\ifthenelse{\boolean{pdflatex}}
{\vspace*{-2.7cm}\mbox{\!\!\!\includegraphics[width=.14\textwidth]{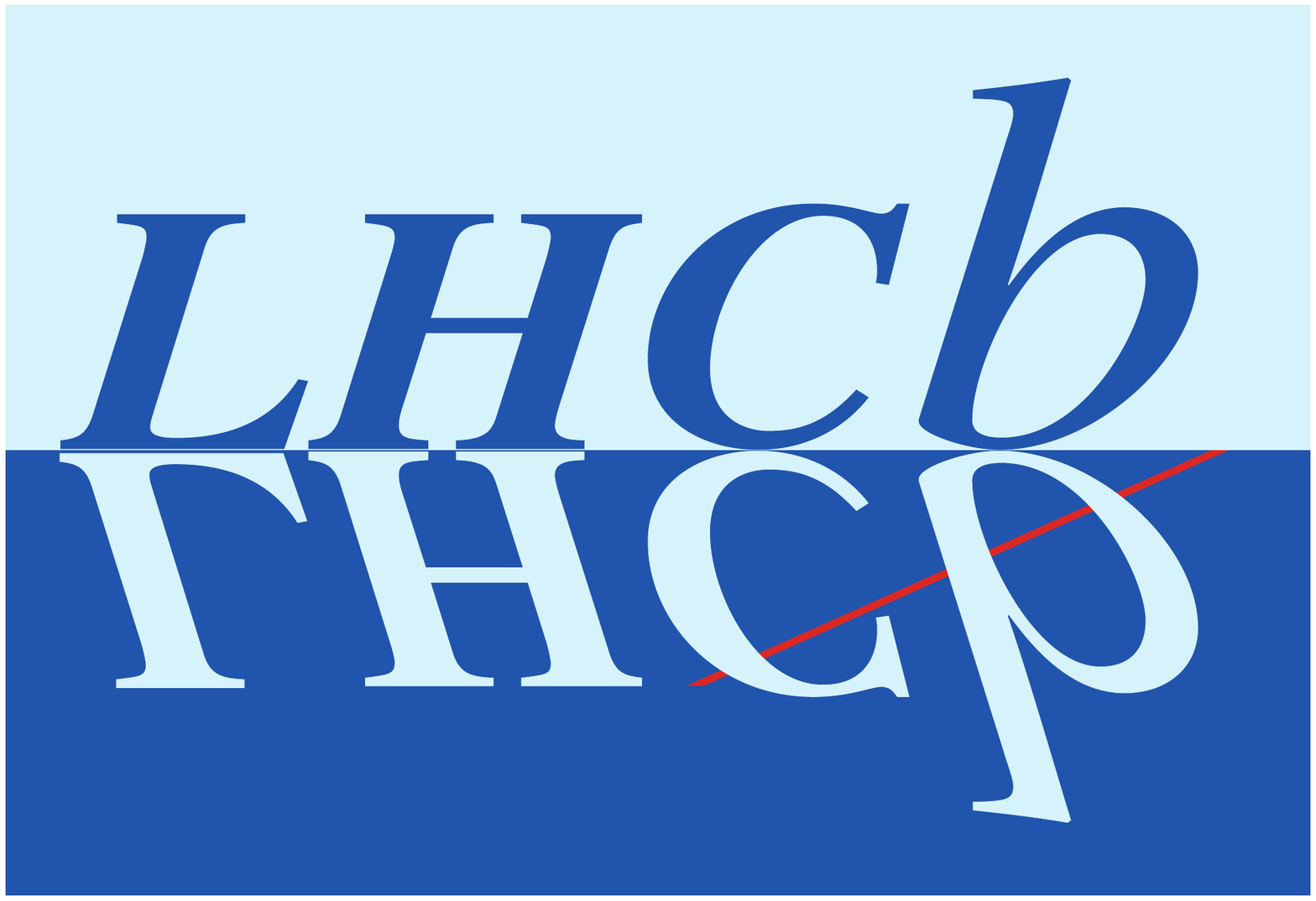}} & &}%
{\vspace*{-1.2cm}\mbox{\!\!\!\includegraphics[width=.12\textwidth]{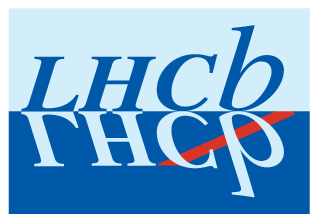}} & &}%
\\
 & & CERN-PH-EP-2013-083 \\  
 & & LHCb-PAPER-2013-013 \\  
 & & May 22, 2013 \\
 & & \\
\end{tabular*}

\vspace*{4.0cm}

{\bf\boldmath\huge
\begin{center}
  Search for the rare decay \dmumu
\end{center}
}

\vspace*{2.0cm}

\begin{center}
The LHCb collaboration\footnote{Authors are listed on the following pages.}
\end{center}

\vspace{\fill}

\begin{abstract}
  \noindent
A search for the rare decay \dmumu  is performed using a data sample, corresponding to an
integrated luminosity of  0.9~\invfb, of 
  $pp$ collisions  collected at
a centre-of-mass energy of  7 TeV  by the LHCb experiment. The observed number of events is consistent with the
background expectations and  corresponds to an  upper limit of
\mbox{${\cal B}(\dmumu) < 6.2 \: (7.6)  \times 10^{-9}$} at \mbox{90\% } \mbox{(95\%)} confidence level.
 This result represents an improvement of more than  a factor twenty with respect to previous measurements.
 \end{abstract}
\vspace*{2.0cm}

\begin{center}
  Submitted to Phys.~Lett.~B 
\end{center}

\vspace{\fill}

{\footnotesize
\centerline{\copyright~CERN on behalf of the \lhcb collaboration, license \href{http://creativecommons.org/licenses/by/3.0/}{CC-BY-3.0}.}}
\vspace*{2mm}

\end{titlepage}


\newpage
\setcounter{page}{2}
\mbox{~}
\newpage

\centerline{\large\bf LHCb collaboration}
\begin{flushleft}
\small
R.~Aaij$^{40}$, 
C.~Abellan~Beteta$^{35,n}$, 
B.~Adeva$^{36}$, 
M.~Adinolfi$^{45}$, 
C.~Adrover$^{6}$, 
A.~Affolder$^{51}$, 
Z.~Ajaltouni$^{5}$, 
J.~Albrecht$^{9}$, 
F.~Alessio$^{37}$, 
M.~Alexander$^{50}$
% ...

{\footnotesize \it

% ...
$ ^{5}$Clermont Universit\'{e}, Universit\'{e} Blaise Pascal, CNRS/IN2P3, LPC, Clermont-Ferrand, France\\
$ ^{6}$CPPM, Aix-Marseille Universit\'{e}, CNRS/IN2P3, Marseille, France\\
% ...
$ ^{9}$Fakult\"{a}t Physik, Technische Universit\"{a}t Dortmund, Dortmund, Germany\\
% ...
$ ^{35}$Universitat de Barcelona, Barcelona, Spain\\
$ ^{36}$Universidad de Santiago de Compostela, Santiago de Compostela, Spain\\
$ ^{37}$European Organization for Nuclear Research (CERN), Geneva, Switzerland\\
% ...
$ ^{40}$Nikhef National Institute for Subatomic Physics, Amsterdam, The Netherlands\\
% ...
$ ^{45}$H.H. Wills Physics Laboratory, University of Bristol, Bristol, United Kingdom\\
% ...
$ ^{50}$School of Physics and Astronomy, University of Glasgow, Glasgow, United Kingdom\\
$ ^{51}$Oliver Lodge Laboratory, University of Liverpool, Liverpool, United Kingdom\\
% ...
$ ^{n}$LIFAELS, La Salle, Universitat Ramon Llull, Barcelona, Spain\\
% ...
}
\end{flushleft}

\cleardoublepage


\renewcommand{\thefootnote}{\arabic{footnote}}
\setcounter{footnote}{0}



\pagestyle{plain} 
\setcounter{page}{1}
\pagenumbering{arabic}


%

\section{Introduction}\label{sec:Introduction}

Flavour-changing neutral current (FCNC) processes are highly suppressed in the
Standard Model (SM) since they are  only  allowed  at loop level and are
affected  by   Glashow-Iliopoulos-Maiani (GIM) suppression~\cite{Glashow:1970gm}. 
They have been extensively studied in processes that involve $K$ and $B$ mesons. 
In  $D$ meson decays, FCNC processes are even more  suppressed by
the GIM mechanism, due to the absence of a high-mass down-type quark.
The \dmumu decay is very rare in the SM because of 
additional helicity suppression. The short distance perturbative  contribution to
the   branching fraction
($\mathcal{B}$) is of the order of 10$^{-18}$ while  
the long distance non-perturbative contribution, dominated by   the two-photon
intermediate state,  is about $2.7 \times 10^{-5}\times {\cal B} (D^0 \to \gamma
\gamma)$~\cite{Burdman:2001tf}. The current upper
limit on ${\cal B} (D^0 \to \gamma \gamma)$ of $2.2 \times 10^{-6}$ at
90\% confidence level (CL)~\cite{Lees:2011qz} translates into an upper
bound for the SM prediction 
for \brmumu of  about $6 \times 10^{-11}$. Given the current upper
limit on \brmumu of $1.4 \times
10^{-7} $ at 90\% CL~\cite{Petric:2010yt},  there is therefore more than  three orders of
magnitude in \brmumu  to be explored before reaching the sensitivity
of  the theoretical prediction. 

Different types of beyond the Standard Model (BSM)  physics could  contribute to  \dmumu decays and
some  could give enhancements with respect  to the short distance
SM prediction of several orders of magnitude. These include   
$R$-parity violating 
models~\cite{Burdman:2001tf,
Burdman:2003rs} and models with Randall-Sundrum warped extra
dimensions~\cite{Paul:2012ab}, with predictions for \mbox{\brmumu} up
to a few times $10^{-10}$.
In general, searches for BSM physics  in charm FCNC processes  are
complementary to those in the $B$ and $K$ sector, since they  provide 
unique access to  up-type dynamics, the charm  being the only
up-type quark undergoing 
flavour oscillations. 

In this Letter, the search for the \dmumu  decay is performed using
\mbox{\dstdmumu} decays, with the $D^{\ast +}$ produced directly at
a $pp$ collision primary \mbox{vertex (PV)}. The inclusion of charge
conjugated processes is implied
throughout the paper.

The data samples used in this analysis were collected during the year
2011  in $pp$ collisions at  a centre-of-mass energy of 7~TeV and correspond
to an integrated luminosity of about 0.9~\invfb. 




\section{Detector and simulation}
\label{sec:Detector}

The \lhcb detector~\cite{Alves:2008zz} is a single-arm forward
spectrometer covering the pseudorapidity range $2<\eta <5$, designed
for the study of particles containing \bquark or \cquark quarks. The
detector includes a high precision tracking system consisting of a
silicon-strip vertex detector surrounding the $pp$ interaction region,
a large-area silicon-strip detector located upstream of a dipole
magnet with a bending power of about $4{\rm\,Tm}$, and three stations
of silicon-strip detectors and straw drift tubes placed
downstream. The combined tracking system provides  momentum
measurement with relative uncertainty that varies from 0.4\% at 5\gevc to 0.6\% at 100\gevc,
and an impact parameter (IP) resolution of 20\mum for tracks with high
transverse momentum. Charged hadrons are identified using two
ring-imaging Cherenkov (RICH) detectors. Photon, electron and hadron
candidates are identified by a calorimeter system consisting of
scintillating-pad and pre-shower detectors, an electromagnetic
calorimeter and a hadronic calorimeter. Muons are identified by a
system composed of alternating layers of iron and multi-wire
proportional chambers. The trigger consists of a hardware stage, based
on information from the calorimeters and muon systems, followed by a
software stage that applies a full event reconstruction~\cite{LHCb-DP-2012-004}.


Events are triggered  and offline-selected in a way that is similar 
for the signal channel \mbox{\dstdmumu}, the normalisation  channel \mbox{\dstdpipi}, and the
control channels  \Jmumu, \mbox{\dstdkpi}, and   \mbox{\dkpi} selected
without  the  $D^{\ast}$  requirement.

All events are triggered at the  hardware stage by requiring one muon  with  transverse
momentum \mbox{\pt$>1.5$ \gevc}  or
two muons with \mbox{$\sqrt{\pt_1 \times \pt_2}> 1.3$ \gevc.} 
Decay channels with muons in the final state, 
\mbox{\dstdmumu}  and \Jmumu, are required to have one of the decay
particles having triggered the event. Channels with only hadrons
in the final state,  \mbox{\dstdkpi}, \mbox{\dstdpipi} and
\mbox{\dkpi},  are  required to be triggered   by particles other than
those forming the candidate decay, called {\it spectator particles}
in the following.

Exceptions to this trigger scheme are made for \Jmumu events, when used to determine the
 trigger efficiency,  and  
 \mbox{\dstdkpi} events, when used to determine the probability of hadron to muon
 misidentification, as described in Sections~\ref{sec:normalisation}
 and~\ref{sec:misid}, respectively. 

The software trigger   selects events, for muonic
final states,  with either one track
identified as a muon with $\pt > 1.0$ \gevc and impact
parameter  with respect to the PV larger than
 0.1~mm,  or with two oppositely-charged tracks identified as
muons, that  form a vertex and have  an invariant mass \mM$> 1$
\gevcc.   For hadronic final states, it selects events with  at least
one track with \mbox{$\pt > 1.7$ \gevc} and  \mbox{$\chi^2_{\rm IP} > 16$}, where $\chi^2_{\rm IP}$  is the
difference between the $\chi^2$ of the PV built with and
without the considered track.  

In a second stage, the software trigger uses algorithms that  reconstruct 
two-body $D^0$ decays using exactly the same  criteria as  the offline
selection  for signal and control samples. 
In the software trigger, all selected  final states are required to have one of
the decay particles having triggered the event. 
Both \mbox{\dstdpipi}
and \mbox{\dstdkpi} candidate events are prescaled 
to comply with the bandwidth requirements of the experiment.  

 Simulation samples are used for determining the relative efficiencies
 between the signal and the normalization modes: $pp$ collisions are generated using
\pythia~6.4~\cite{Sjostrand:2006za} with a specific \lhcb
configuration~\cite{LHCb-PROC-2010-056}; decays of hadronic particles
are described by \evtgen~\cite{Lange:2001uf} in which final state
radiation is generated using \photos~\cite{Golonka:2005pn} and the
interaction of the generated particles with the detector and its
response are implemented using the \geant
toolkit~\cite{Allison:2006ve, *Agostinelli:2002hh} as described in
Ref.~\cite{LHCb-PROC-2011-006}.

\section{Candidate selection}
\label{sec:selection}

Candidate \dmumu decays are reconstructed in \dstdpi decays.
The two $D^{0}$
daughter tracks are 
required to be of good quality ($\chi^2$ per degree of freedom (ndf)
$<$ 5) and  to be displaced
with respect to the closest PV, with  $\chi^2_{\rm IP}$
larger than 3 and 8 and  \pt larger than 
750 \mevc and
1100 \mevc. The $D^0$  secondary vertex (SV) is required to
be of good quality ($\chi^2_{\rm SV}<$ 10) and clearly separated from the
PV in the forward direction (vertex separation
\mbox{$\chi^2 > 20$}). 
When more than one PV per event is reconstructed, the one
that gives the minimum  $\chi^2_{\rm IP}$  for the candidate
is chosen. The $D^0$ candidate has to point to the PV ($\chi^2_{\rm IP}<$ 15 and cos$(\theta_{\rm P})>0.9997$, where
$\theta_{\rm P}$ is the angle between the $D^0$ momentum in the laboratory frame and
the direction defined by the PV and SV) and have
\mbox{\pt$>$~1800\mevc} . The same selection is also applied to  \Jmumu
candidates, which are used for validating the muon identification  and
 trigger efficiency derived from the simulation.  
Candidate  $D^0$ (\jpsi) mesons are selected if their decay products
have an invariant mass in the region of the known  $D^0$ (\jpsi)  mass.

An additional selection requirement, not applied
at the trigger stage,  is that the bachelor $\pi^+$ of the \dstdpi
decay has 
$\chi^2_{\rm IP}<$ 10, \pt$>$ 110 \mevc  and 
 is  constrained to the PV using  a Kalman
filter (KF)~\cite{Hulsbergen:2005pu}. This provides 
an improved resolution for the mass difference  between the $D^{\ast
  +}$ and $D^0$ candidates,  \mbox{\deltamh$\equiv$ \deltamhex},
where $h=\mu,\pi$ and $h^{\prime}=K,\mu,\pi$. 
Candidates are  selected  with a mass difference value around  145.5\mevcc.

After the selection, the background of the signal channel
has two main sources: a peaking background due to two- and three-body $D^0$ decays, with
one or two hadrons misidentified as muons, and 
 combinatorial background due to semileptonic decays of beauty  and
charm hadrons. 
The former is  reduced with tight particle identification
criteria  while
the latter is reduced applying  a multivariate selection. 

The muon candidates in the $D^0$ decay  are required to have
associated muon chamber hits that are not shared with any other track in the
event.
A cut  on a combined particle identification likelihood, aimed at
separating muons from other particle
species~\cite{1748-0221-8-02-P02022}, is  applied.  This likelihood combines 
information about track-hit matching in the  muon chambers, energy
associated to the track in the calorimeters and particle
identification in the RICH detectors. 
In order to explicitly veto kaons, thereby suppressing  backgrounds
such as \dstdkmunu decays, an additional cut on a  particle
identification likelihood aimed at separating kaons from other
particle species~\cite{2012arXiv1211.6759A} is applied. The remaining
dominant source of pion to muon misidentification is due to 
pion decays in flight.
 
A boosted decision tree  (BDT)~\cite{Breiman}, with  the Ada\-Boost
algorithm~\cite{AdaBoost},  provides a multivariate discriminant and
is based on the
following variables: $\chi^2_{\rm KF}$ of the 
 constrained  fit, $\chi^2_{\rm IP}$
of the $D^0$  vertex,   $D^0$  pointing angle
$\theta_{\rm P}$,
minimum \pt and  $\chi^2_{\rm IP}$ of the two muons, positively-charged muon angle in the $D^0$ rest frame
with respect to the $D^0$ flight direction and  $D^0$ angle in the $D^{\ast +}$
rest frame with respect to the $D^{\ast +}$ flight direction. 
  The BDT  training makes use of  \mbox{\dstdmumu} simulated events
 and \mM sideband data  (\mbox{1810$-$1830 \mevcc} and \mbox{1885$-$1930 \mevcc}); the data sample for
the training consists of a separate sample of 
80\invpb,  satisfying  the same selection criteria,  and  is not  used in the search for
the \dmumu decay.
The absence of   correlation between   \mM  and the BDT
output variable 
is explicitly checked  using  data selected
with the cuts   \mbox{\deltamM$>147$ \mevcc} and \mbox{\mM$>1840$ \mevcc.}
The cut value on the BDT output variable is chosen in order to
achieve the best expected limit  on \brmumu, based on  simulated
pseudo-experiments (see Section~\ref{sec:fit}) assuming no
 signal,  and has  an efficiency of 74\% on the signal while providing a 
 reduction of more than a factor  of 
three for the combinatorial background.

\section{Normalisation }
\label{sec:normalisation}

The \dmumu branching fraction is obtained from
\be
\mathcal{B}(\dmumu) = \frac{N_{\peddstdmumu}}{N_{\peddstdpipi}} \times
\frac{\varepsilon_{\pi\pi}}{\varepsilon_{\mu\mu}}\times \mathcal{B}(\dpipi) = \alpha \times N_{\peddstdmumu}
\label{eq:br_norm}
\ee
\noindent using the decay \dstdpipi as a normalisation mode, where
$\alpha$ is the single event sensitivity, $N_{ \peddstdpipi(\peddstdmumu)} $ are
the yields and 
$\varepsilon_{\pi\pi(\mu\mu)}$  the total
efficiencies for \dstdpipi (\dstdmumu)
decays. 
In this section  the various contributions to $\alpha$ are determined. 


The trigger efficiencies for the signal and normalisation channels
are calculated 
using  the simulation and corrected  using data driven methods. 
To cross-check the signal trigger efficiency, \Jmumu  events are
selected in both data and simulation and  triggered
using spectator particles; consistency is observed  within the relative statistical uncertainty of 2.7\%.
To cross-check the trigger efficiency for the normalisation channel,  the
trigger efficiency of the  \dkpi decay is  measured in a sub-sample of
data  taken with very loose trigger requirements.
A correction factor for the trigger efficiency of \dstdpipi events as
derived from the simulation is obtained, with  a 
systematic uncertainty of 4.9\%, mostly arising from the statistical
uncertainty on the correction itself.
The  trigger efficiencies are found to be (86.4$\pm
$2.4)\%  and (3.3$\pm $0.2)\% for  \dstdmumu and  \dstdpipi, respectively. The
low efficiency of the hadronic channel comes from the requirement of   
the event being  triggered   by spectator particles.

The muon identification efficiency is also measured with simulated events and  validated 
with  a sample of \Jmumu decays triggered   by spectator particles. The efficiency of the requirement of
having associated hits in the muon chambers is  determined using
\Jmumu decays  selected without muon 
identification requirements on one of the two tracks. The efficiency correction
of the combined particle identification likelihood is determined
comparing \Jmumu decays selected  with the above muon identification
criteria  and  in a kinematic of 
region of mean transverse momentum (\mbox{$\langle \pt(\mu)\rangle < 1$\gevc}) and
opening angle of the two muons in the plane transverse to
the beam ($\Delta \phi< 1$ rad) which is similar to that of the
\dstdmumu decays.
Good agreement between data and simulation is observed within 2.6\%,
which is assigned as a systematic uncertainty.

Acceptance, reconstruction and selection efficiencies for the signal and normalisation channels are measured
using  simulated events. In order to estimate discrepancies
between data and  simulation, the \mbox{\dstdkpi} channel, which benefits from high yield and  low background, is used.
Small deviations from the simulation shapes are observed  in the $D^0$ daughter impact parameter,
momentum and transverse momentum distributions.
  Since, for the  branching fraction measurement, only the efficiency ratio matters,
  any systematic uncertainty related to these quantities cancels at first order. Indeed, varying
  the cuts, it is verified that  the ratio of selection
  efficiencies changes by a negligible amount compared to  other
  systematic uncertainties on $\alpha$.
The effect of interactions of the decay products with the
detector material, different for
muons and pions, results in an  additional systematic uncertainty
of 3\% per track~\cite{Stone}. 
The  selection and reconstruction efficiency ratio between the normalisation  and 
signal channel
is found to be 1.17$\pm$0.08, with the deviation from unity  mostly
coming from the muon identification efficiency.

The yield extraction for the normalisation channel  is performed with an 
unbinned extended maximum likelihood fit to the two-dimensional distribution 
 of \deltamP  and \mP. 
The probability density functions (PDFs) that parametrize  the
\deltamP  distribution are a double Gaussian shape with common mean for the
signal and the  parametric function
\be
f_\Delta(\deltamP,a,b,c) = \left( 1-e^{-(\Delta m_{\pi^{+}\pi^{-}}   -\Delta m_0)/c}\right)
\times \left(\frac{\deltamP}{\Delta m_0}\right)^{a} + b\times
  \left(\frac{\deltamP}{\Delta m_0}-1\right) 
\ee
for the combinatorial
background, where \mbox{$\Delta
  m_0=139.6 \mevcc$} and $a$, $b$ and $c$ are fit parameters.  For the
\dstdpipi fit only  the $c$ parameter is varied and $a$ and $b$  are set to
0.  The \mP
distribution is parametrized with a Crystal Ball (CB)~\cite{Skwarnicki:1986xj} function
for the signal and  a single exponential shape for the combinatorial
background. The CB  is a four-parameter function consisting  of a Gaussian core, of mean $\mu$ and
width $\sigma$,  and a 
power-law low-end tail with negative slope
$n$, below a  threshold, defined by the $\omega$ parameter, at the
value $\mu-\omega\times \sigma$. 
A small background  component due to the random
association of a \dpipi decay with a pion from the PV is also added to the fit,
with the same PDF as the \dstdpipi for the \mP  distribution, and
 the same $f_\Delta$ function as the combinatorial
background for the \deltamP distribution.  For the \dstdkpi component,
a Gaussian PDF  for the \deltamP  distribution  and a single
exponential function   for the \mP distribution are used.

\begin{figure}[!htbp]
\begin{center}
\begin{overpic}[width
= 0.69\textwidth]{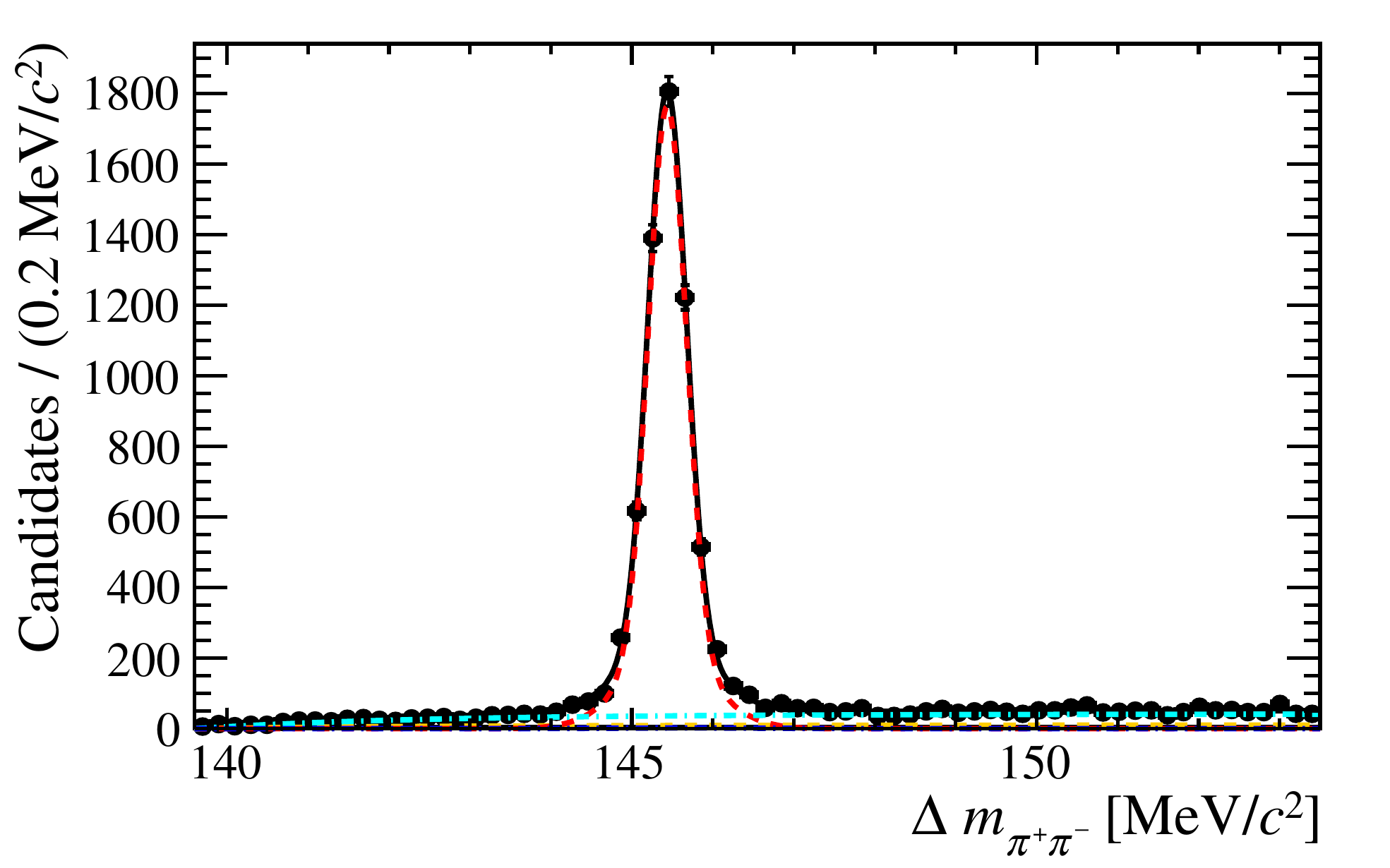}
\put(20,50){(a)}
\put(80,50){LHCb}
\end{overpic} 
\begin{overpic}[width = 
0.69\textwidth]{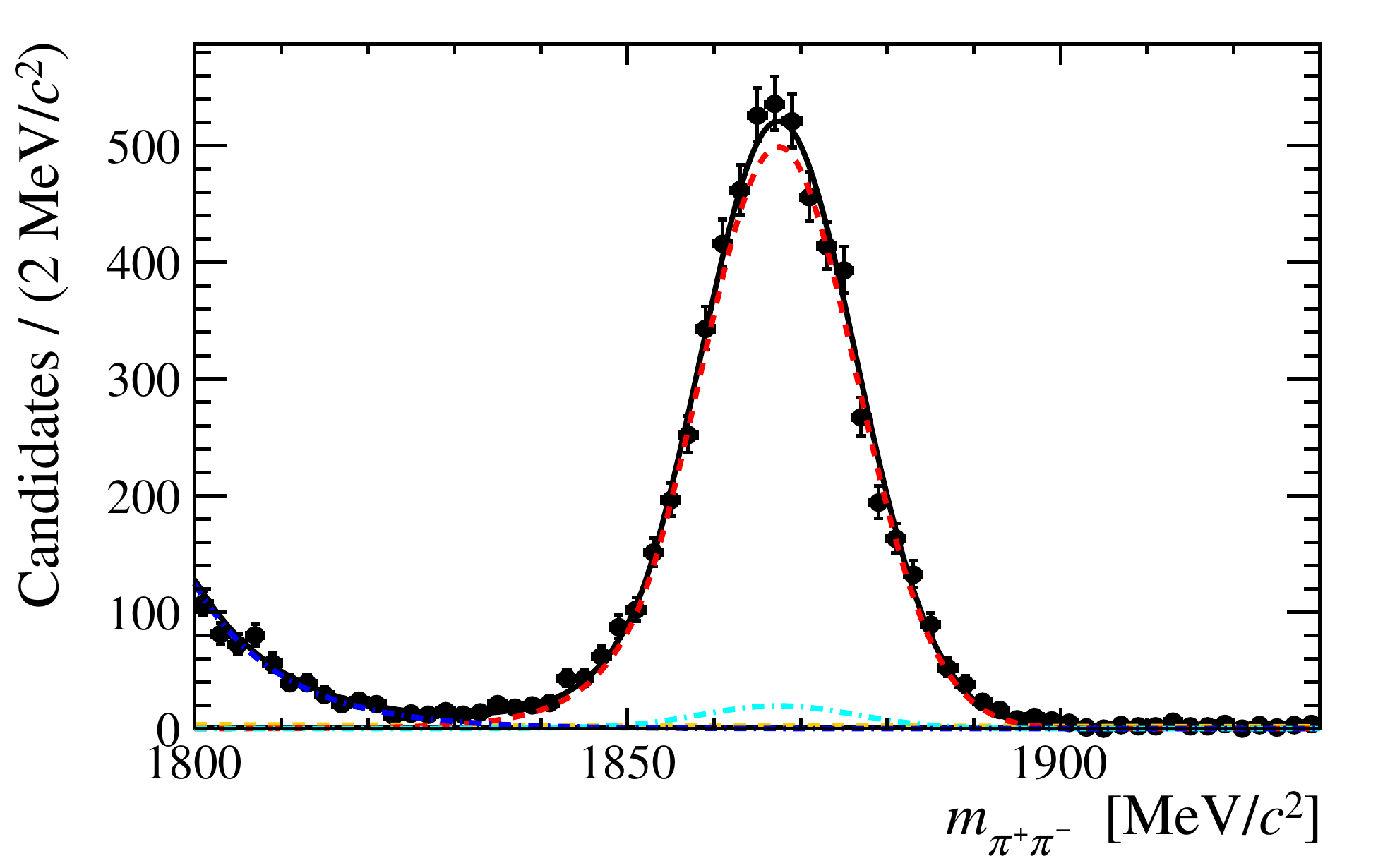}
\put(20,50){(b)}
\put(80,50){LHCb}
\end{overpic}
\caption{\small (a) Invariant mass difference \deltamP,
  with \mP in the range 1840$-$1885\til\mevcc, and  (b) invariant mass  \mP, with \deltamP  in the range
144$-$147\til \mevcc,  for 
\mbox{\dstdpipi} candidates in data. The  projections of the two-dimensional
unbinned extended maximum likelihood fit  are overlaid.  The curves represent   the total
 (solid black),  \dstdpipi  (dashed red), the  untagged
\dpipi (dash-dotted cyan), the combinatorial background (dashed yellow)
and the \dstdkpi (dash-dotted blue) contributions.
The  \dstdpipi candidates are prescaled at the software trigger stage by a factor 0.03. }\label{fig:dpipi_mass_fits}
\end{center}
\end{figure}

Figure~\ref{fig:dpipi_mass_fits} shows the \deltamP
 and  the \mP distributions for \mbox{\dstdpipi} from
 which a total yield of 6201$\pm $88 decays is estimated. 
The total  uncertainty on the yield is  dominated by 
statistics. It has been  verified that alternative PDF
parametrizations, such as modifications  of the $f_\Delta$ function, do
not lead to significant changes in the extracted \dstdpipi yields. 

A single event sensitivity  of $(3.00  \pm 0.27)
\times 10^{-10}$ is  obtained.  The systematic contributions
to $\alpha$, together with the total uncertainty, obtained by summing
in quadrature the individual contributions,  are summarized in Table~\ref{tab:sys}.

\begin{table}[!htbp]
\caption{\small  Contributions to the  systematic uncertainty of the single event sensitivity $\alpha$.} 
\label{tab:sys}
\begin{center}
\begin{tabular}{lc}
  Source & Relative uncertainty (\%) \\ 
\midrule
Material interactions &  6.0 \\
Muon identification efficiency & 2.6 \\
Hadronic trigger efficiency & 4.9 \\
Muon trigger efficiency & 2.7 \\
$\mathcal{B}(\dpipi)$~\cite{Nakamura:2010zzi} &  1.9  \\ 
\midrule
Total systematic uncertainty &  8.8 \\
\end{tabular}
\end{center}
\end{table}

\section{Background yields from { \boldmath \dstdpipi}
decays}
\label{sec:misid}

Due to the similar topology to the signal decay channel and the small difference between the pion and the muon mass, only the
\dstdpipi decay can significantly contribute  as peaking background in
both the \mM and \deltamM distributions  when the two pions are misidentified as muons. 

The yield of  misidentified \mbox{\dstdpipi} decays, $N_{\peddstdpipimumu}$, is obtained from
the yield of selected \dstdpipi events, $N_{ \peddstdpipi} $,  as 
\be
N_{\peddstdpipimumu} = N_{\peddstdpipi} \times 
\frac{ \varepsilon_{\pi\pi\to\mu\mu}}{\varepsilon_{\pi\pi}}
\label{form:misid}
\ee
where $\varepsilon_{\pi\pi\to\mu\mu}$ is the total efficiency for \dstdpipi
events with both pions misidentified as muons. Both the efficiencies in the numerator and denominator 
are obtained from the 
simulation of a very large \dstdpipi event sample and corrections are
applied using data driven methods. 
The main systematic uncertainties in Eq.\til\ref{form:misid} come from the trigger efficiency
for \dstdpipi events, as discussed in
Section\til\ref{sec:normalisation}, and the  misidentification
probability, which  is cross-checked with data using  \dstdkpi events.

\begin{figure}[!htbp]
\begin{center}
\begin{overpic}[width =
0.69\textwidth]{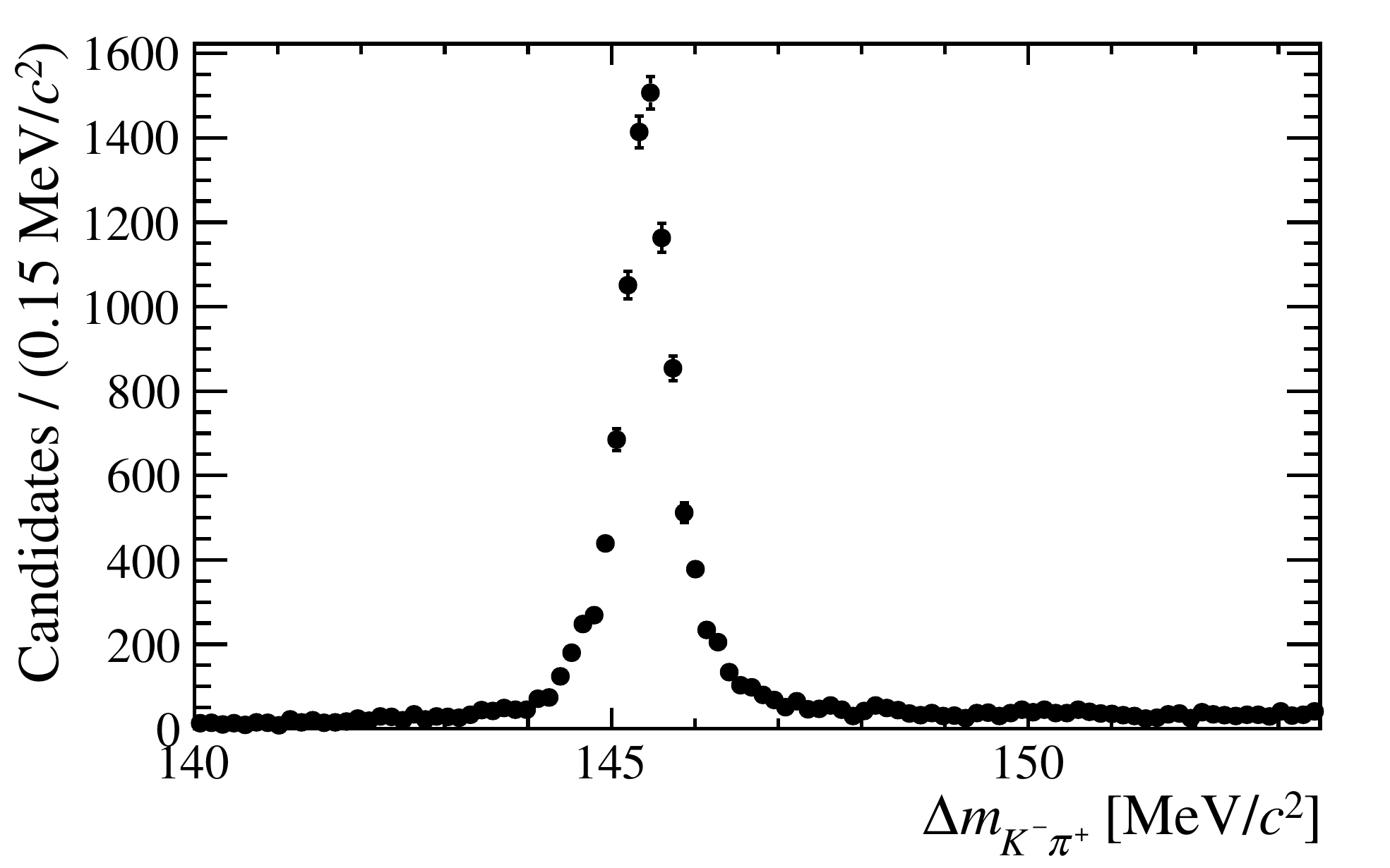}
\put(20,50){(a)}
\put(80,50){LHCb}
\end{overpic} 
\begin{overpic}[width = 
0.69\textwidth]{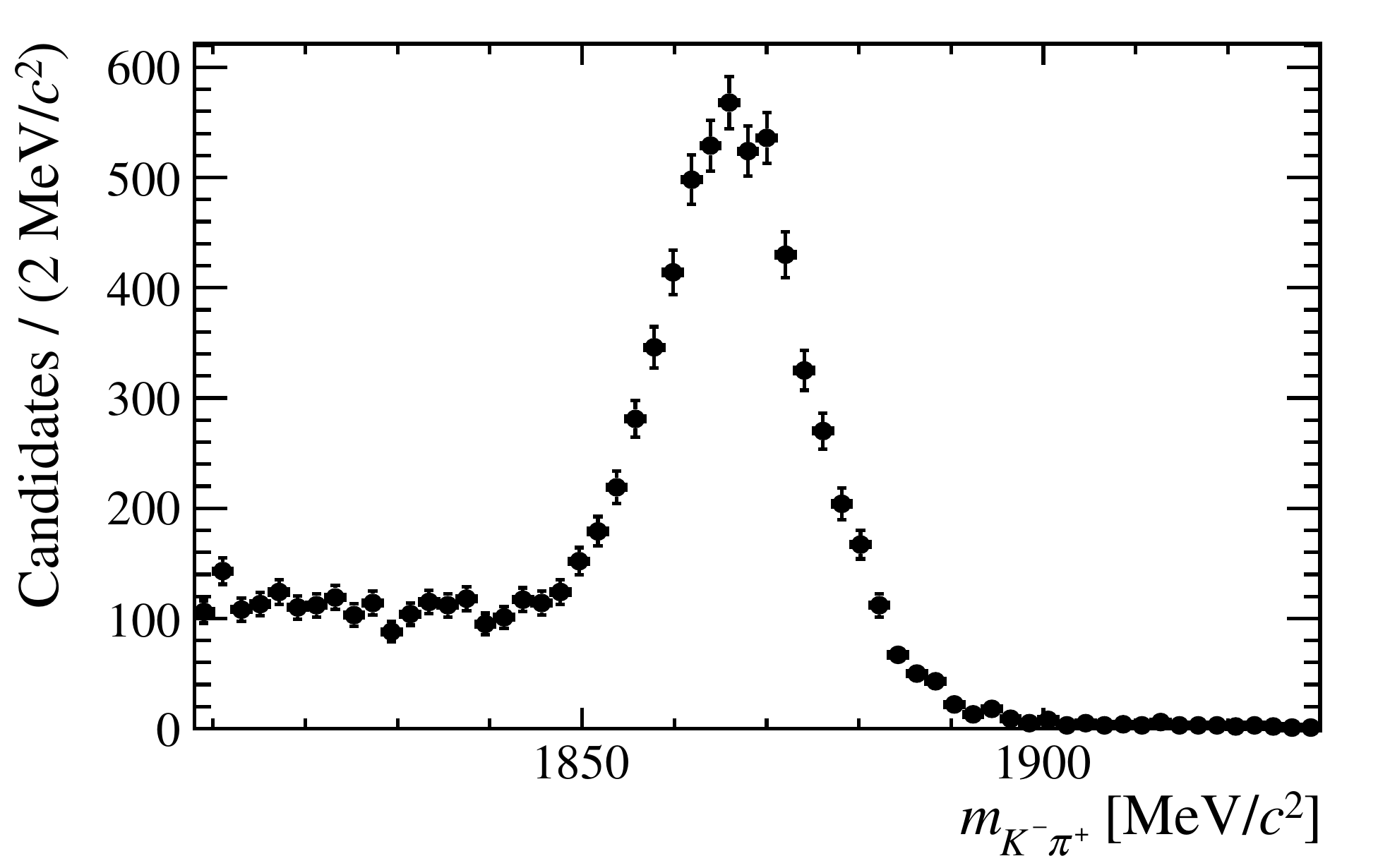}
\put(20,50){(b)}
\put(80,50){LHCb}
\end{overpic}
\caption{ \small  (a) Invariant mass difference \deltamK  and (b)
  invariant mass \mK  distributions in data for \dstdkpi candidates, 
 with muon identification applied to the pion.
The events are  triggered by  spectator particles with respect to the $D^0$
 daughter  pion. The  \dstdkpi candidates are prescaled at the software trigger stage by a factor 0.15.}
\end{center}
\label{fig:misid_masses_kpi}
\end{figure}

The invariant mass difference  \deltamK and invariant mass \mK
distributions in data for \dstdkpi candidates,  with muon
identification applied to the pion and triggered by  spectator particles with respect to the $D^0$
 daughter  pion, are shown in Fig.~\ref{fig:misid_masses_kpi}.
These distributions show some  remarkable  differences compared to  those of   
Fig.~\ref{fig:dpipi_mass_fits}:  a  tail
appears on the left of the peak in  the \mK distribution and the \deltamK and \mK distributions
display a broader
mass distribution. As
the simulation shows, the left-hand tail in the \mK distribution
comes  from two effects of comparable size, a   low mass tail of the \dstdkpi
decays  due to 
the momentum loss in the   pion
decay 
and a high mass tail of  \mbox{\dstdkmunu} decays, the latter also
contributing to the broadening of
the mass resolution of the \deltamK distribution. To suppress the
background from  \mbox{\dstdkmunu} decays, the  measurement of  the  misidentification probability
is performed using only  the candidates on the upper side of the   \mK  peak, taking the ratio of
events with and without the muon identification applied to the pion. A
correction factor of \mbox{1.2 $\pm$ 0.1}, taking into account the event yield in  the pion
decay tail, is  applied.

The single pion to muon 
 misidentification probability  in data is \mbox{$(2.9 \pm
 0.2)\times 10^{-3}$}, with a  ratio of data to simulation  of   $0.88
 \pm 0.15$. This latter value, squared, is 
used as a correction factor for  ${\varepsilon_{\pi\pi\to\mu\mu}}$ as determined from the simulation.
A reweighting of the single misidentification
 probability taking into account the momentum correlation of the two
 $D^0$ daughter pions gives consistent  results within the uncertainties.  It
 is also verified that the small difference in the momentum
 distribution between data and simulation has a negligible  impact on
 the determination of the misidentification probability.

The number of expected doubly-misidentified \dstdpipi
decays in our data sample  \mbox{is $45 \pm 19$}. 

\begin{figure}[!htbp]
\begin{center}
\begin{overpic}[width=
0.69\textwidth]{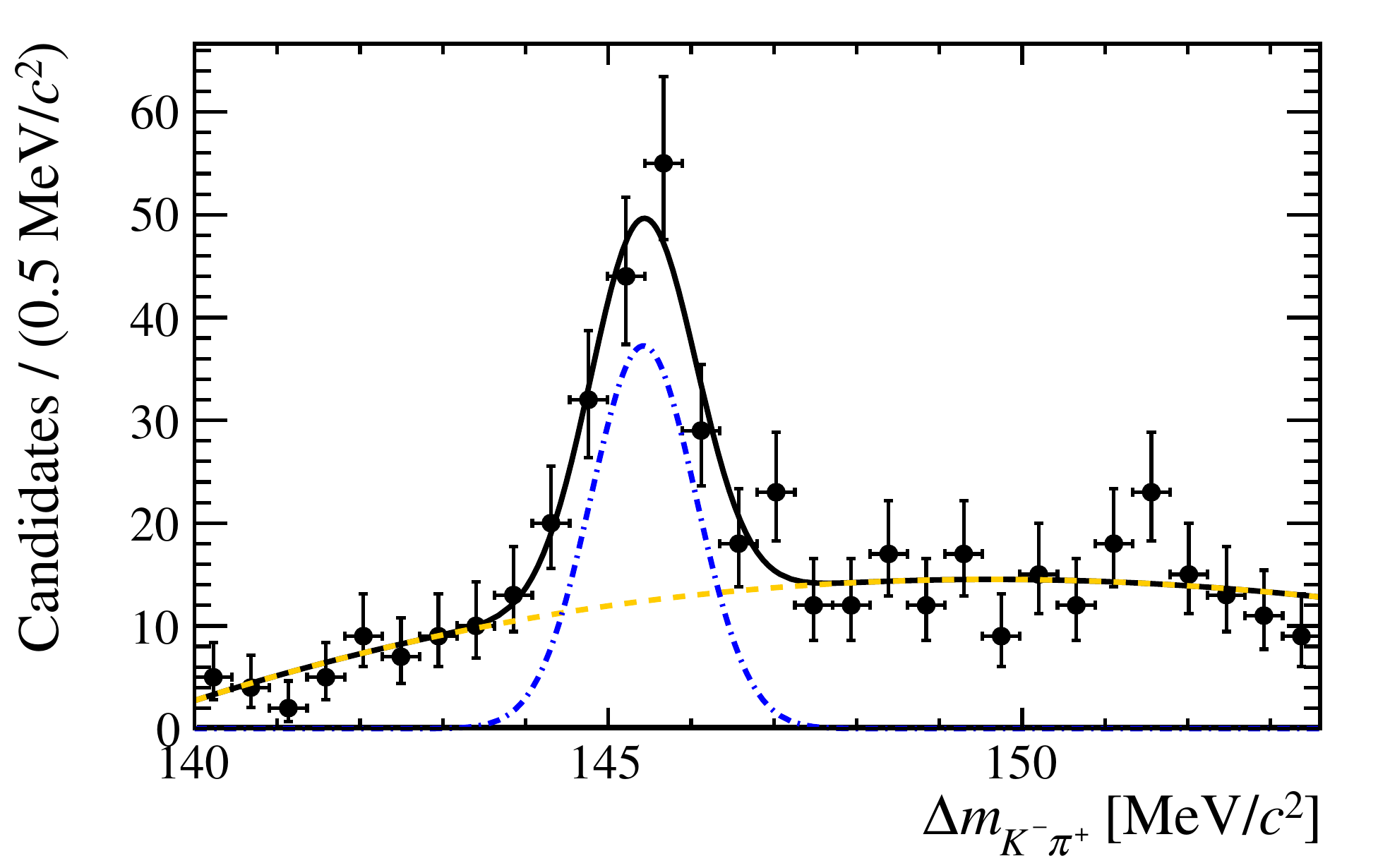} 
\put(80,50){LHCb}
\end{overpic}  
\end{center}
\caption{\small  Invariant mass difference \deltamK  for
  \dstdkpi candidates, with both hadrons misidentified as muons, with
  \mM in the range \mbox{1720$-$1800 \mevcc}. The two muons are reconstructed using  the
$K\pi$ and 
$\pi K$ mass hypotheses.   The result of the  
 unbinned extended maximum likelihood fit is overlaid. The curves represent   the  total
 distribution (solid black), the combinatorial background (dashed yellow)
 and the \dstdkpi (dash-dotted blue) contribution.}
\label{fig:kpimumu}
\end{figure}

The determination  of the number of doubly-misidentified events is
cross-checked  by considering  the observed  number of \dstdkpi candidates with
double misidentification in the lower sideband of the \mM
distribution, extending in the selection down to 1720 \mevcc.
The yield of these events is determined  from an unbinned  extended maximum likelihood fit
to the  \deltamK distribution, where the two muons  are reconstructed using  the
$K\pi$ and 
$\pi K$ mass hypotheses, requiring  \mbox{\mM$<$1800 \mevcc}, as
shown in Fig.~\ref{fig:kpimumu}. The PDFs used for the fit are a
single Gaussian shape for the \dstdkpi events with floating  mean and width and
an $f_\Delta$ function for the background, will all three $a$, $b$ and
$c$ parameters allowed to vary. 
To obtain a prediction for the number of  misidentified
\dstdpipi decays, the number of \dstdkpi candidates   is   multiplied by the ratio of \dpipi to  \dkpi branching
fractions  and by the ratio of pion to
muon and kaon to muon  misidentification probabilities, assuming  factorization of the 
misidentification probabilities of the two $D^0$ daughters.  The kaon to muon
misidentification probability is measured with \dstdkpi 
decays,  triggered by  spectator particles with respect to the kaon,
and is found to be (6.3$\pm$0.6)$\times 10^{-4}$. This very small value is achieved using  the kaon veto  based on the RICH detectors, as described in Section~\ref{sec:selection}.
The estimated  yield of \dstdpipi is   compatible with that obtained
from  the method described above, though with a larger uncertainty.

\section{Results}
\label{sec:fit}

The search for the   \dmumu decay is performed using  an 
 unbinned extended maximum likelihood fit to  the two-dimensional
 distribution of 
\deltamM  and   \mM.
The five different fit components are the 
signal \dstdmumu, the combinatorial background and the background from
 \dstdpipi,
\dstdkpi and 
\dstdpimunu decays.

The PDF shapes are
chosen as detailed in Table~\ref{tab:shapes}.  The parameter input values  are determined from
the  simulation of the  individual  channels, except  for the
combinatorial background, which is assumed to have a smooth
distribution across the whole invariant mass difference \deltamM  and
invariant mass   \mM ranges, as in the
\dstdpipi fit of Section~\ref{sec:normalisation}. The table also
shows the corresponding  fit parameters that  are allowed to vary, both freely
and with Gaussian constraints. Other fit parameters,  not included in the table,
are fixed to the values obtained from the simulation. It is
explicitly checked that the final result is insensitive to the
variation of these  parameters.

The width of the CB function
describing the \dstdpipi background in the \mM distribution  and
 the narrower  width of
   the double Gaussian shape  describing the \dstdpipi
  background in the \deltamM distribution  are corrected for the
 broader mass  distribution observed in data; the widths are  increased by about 40\% in \deltamM
 and 25\% in \mM. The CB slope parameter  is  fixed to the mean
   value obtained  from  simulation. Varying this value
 within its  uncertainty leads to a negligible change in the
 final result. 

The \dstdkpi and  \dstdpimunu yields are normalized
 to the \dstdpipi yields based on their relative branching fractions, on the
 number of generated events and on the pion to muon and kaon to muon
 misidentification probabilities, as measured from  data. To take
 into account discrepancies between data and  simulation for these two latter
 quantities,  a conservative  uncertainty of 50\% and 30\% is assigned,
 respectively. \\
The signal  PDFs are parametrised as in the \dstdpipi fit of
Section~\ref{sec:normalisation} and the shape parameters are fixed to
 the \dstdpipi output fit values. A variation of these parameters
 within their uncertainties 
 give a negligible effect on the final value for \brmumu.

\begin{table}[!htb[]
\caption{\small   PDF components  describing \mM and
  \deltamM  distributions  in the signal and corresponding freely  varying 
   and  Gaussian constrained  fit  parameters. The coefficients of the
  exponential (EXP) function used to describe both the \dstdkpi and  \dstdpimunu
  backgrounds are $\gamma_{K\pi}$ and  $\gamma_{\pi\mu\nu}$ while
  $f_{K\pi}$ and $f_{\pi\mu\nu}$  are the normalisation factors to the
  \dstdpipi events. The symbols ${\langle\deltamM\rangle}_{\eta}$, $\eta=i, j$ and
  $k$  represent  the mean values 
 and  $(\sigma^{\Delta}_1)_{\eta}$  the narrower width  of the
   double Gaussian  (DG) PDF describing \dstdpipi, \dstdkpi and \dstdkmunu
   distributions (for \dstdkpi a single Gaussian (SG) PDF is used). The normalisation for the 
  \dstdpipi event yield  is obtained  from the procedure described in
  Section\til\protect\ref{sec:misid}. The
  function  $f_m$ is a constant. The parameters
  $\omega$, $\mu$ and $\sigma$ of the Crystal Ball function describing the
  \dstdpipi events are described in
  Section\til\protect\ref{sec:normalisation}. }\label{tab:shapes}
\begin{center}
\begin{tabular}{lcccc}
Fit component  & \mM  & \deltamM  &     Free & Constrained    \\
\midrule    
 Combinatorial &  $f_m$ &  $f_\Delta$  & yield,  & \\
 &   &     &  $a$, $b$, $c$    & \\

\dstdpipi &  CB & DG &  & $\alpha$,
$\varepsilon_{\pi\pi\to\mu\mu}$, $\omega$, $\mu$, $\sigma$,
 \\
& & &  & 
${\langle\deltamM\rangle}_i$, $(\sigma^{\Delta}_1)_i$ \\
\dstdkpi &  EXP & SG & & $\gamma_{K\pi}$, $f_{K\pi}$, ${\langle\deltamM\rangle}_j$, $(\sigma^{\Delta})_j$ \\
\dstdpimunu &  EXP & DG &  & $\gamma_{\pi\mu\nu}$,
$f_{\pi\mu\nu}$, ${\langle\deltamM\rangle}_k$, $(\sigma^{\Delta}_1)_k$  \\
\dstdmumu &  CB & DG & yield &   \\
\end{tabular}\end{center}
\end{table}

The  systematic uncertainties related  to both
the normalisation, through $\alpha$, and  the background shapes and
yields,
are included in the fit as Gaussian constraints to the parameters.

After the fit, all constrained  parameters converged to the input
values  within a few percent  but  $\varepsilon_{\pi\pi\to\mu\mu}$
and $\omega$, which changed by about $+16\%$ and $-20\%$, respectively,
though  remaining consistent  with the fit input values, within the uncertainty. 

\begin{figure}[!htbp]
\begin{center}
\begin{overpic}[width=
0.69\textwidth]{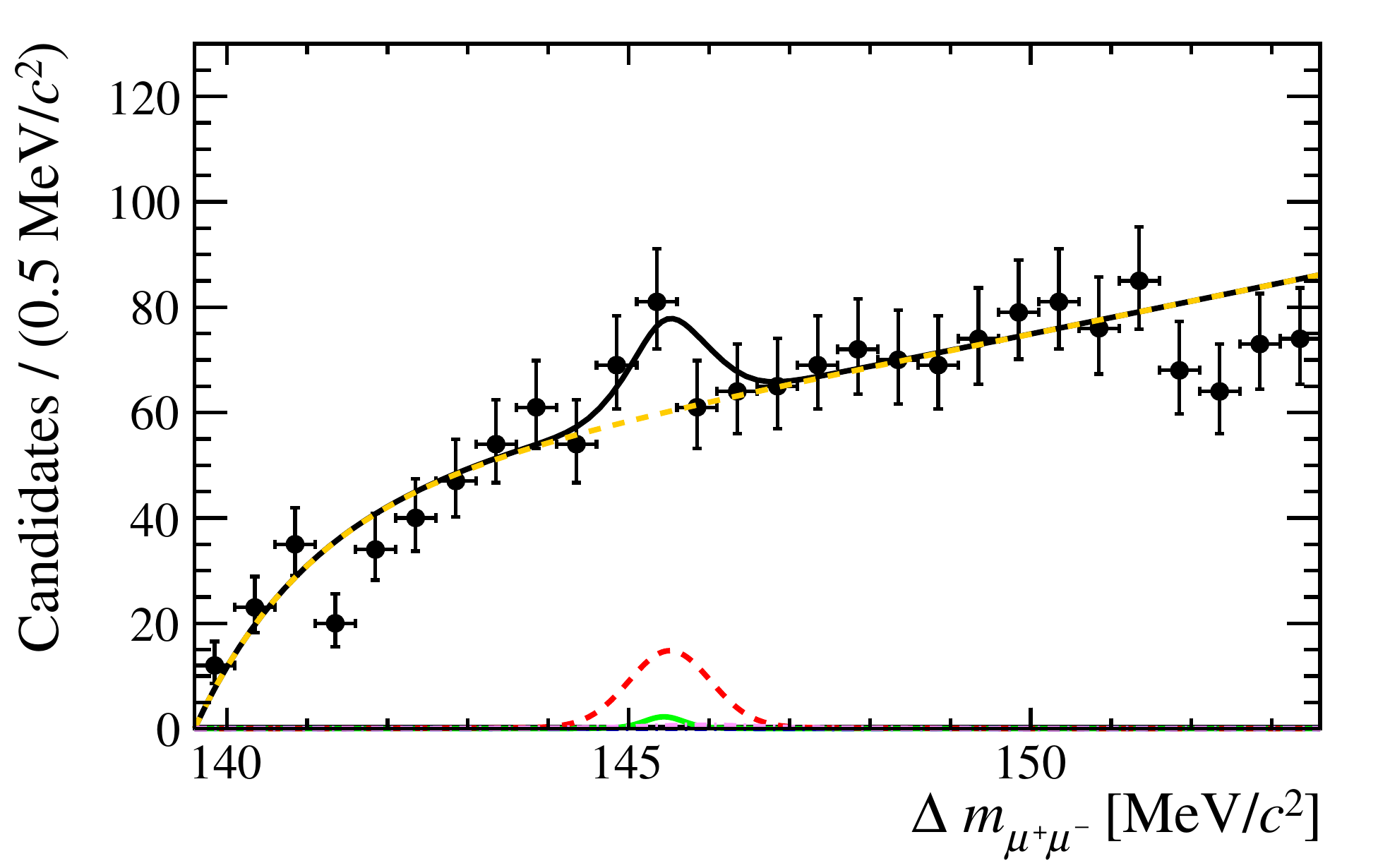}
\put(20,50){(a)}
\put(80,50){LHCb}
\end{overpic}  
\begin{overpic}[width = 0.69
\textwidth]{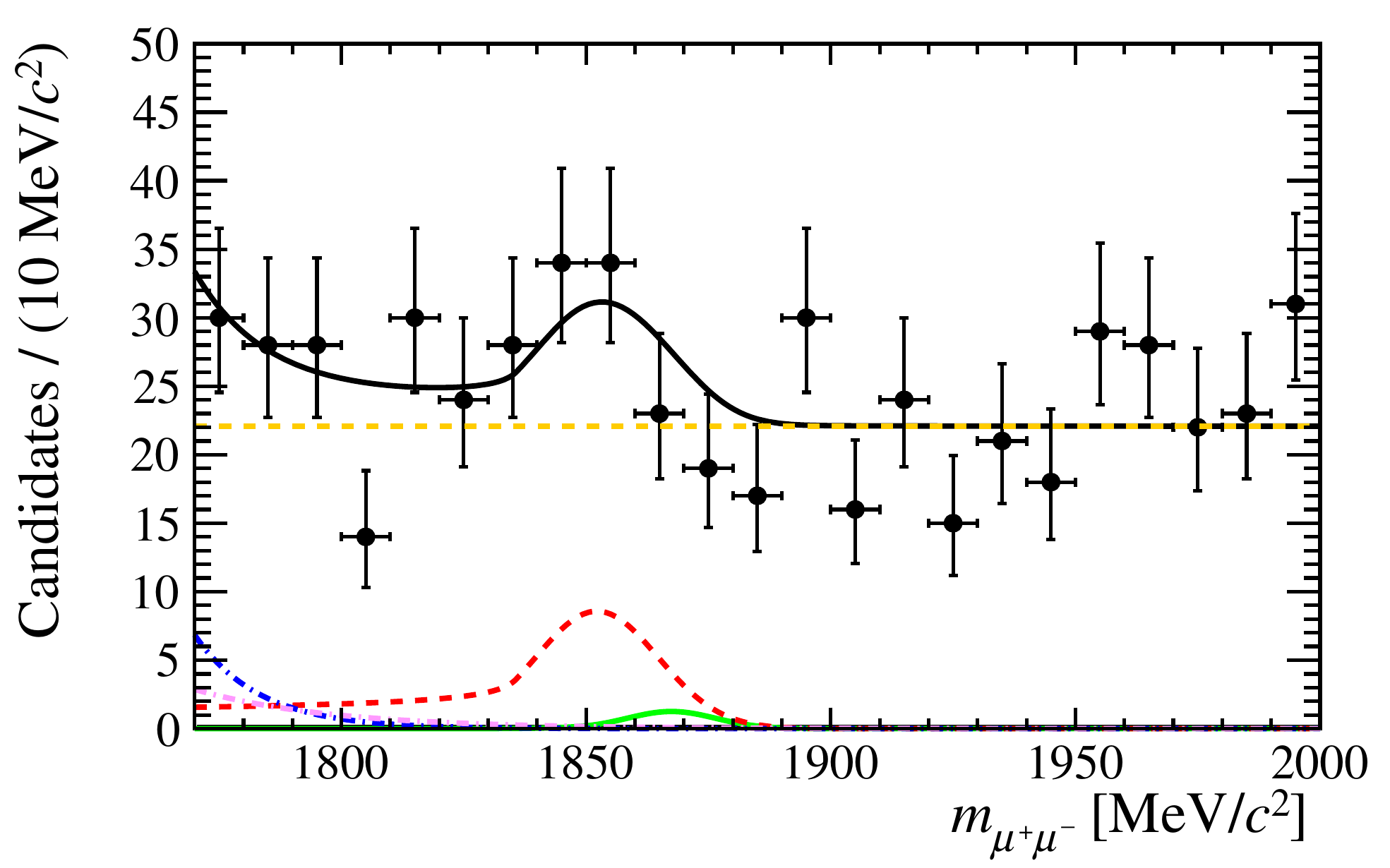}
\put(20,50){(b)}
\put(80,50){LHCb}
\end{overpic}
\end{center}
\caption{\small  (a) Invariant mass difference \deltamM,
  with \mM  in the range 1820$-$1885\til \mevcc and 
  (b) invariant mass \mM, with   \deltamM in the range
144$-$147\til \mevcc for \dstdmumu candidates.  The projections of the  
two-dimensional unbinned extended maximum likelihood fit are overlaid. The curves
represent   the  total distribution (solid black), the \dstdpipi
(dashed red), the combinatorial background (dashed yellow),  the
\dstdkpi (dash-dotted blue), the  \dstdpimunu
(dash-dotted purple) and
the signal \dstdmumu (solid green) contribution. 
}\label{fig:mumu_masses}
\end{figure}

Figure~\ref{fig:mumu_masses} shows the \deltamM and \mM
distributions, together with 
the one-dimensional binned projections of the two-dimensional fit overlaid. The
$\chi^{2}$/ndf of the fit projections are 1.0 and 1.3, corresponding to  probabilities of 44\% and 19\%, respectively.
The data are consistent with the expected backgrounds.
In particular, a residual
contribution from \dstdpipi events  is  visible among the peaking backgrounds.

The  value obtained for the \dmumu branching fraction is  \mbox{$(0.09\pm 0.30) \times 10^{-8}$}. 
Since no significant
excess of signal is observed with respect to the expected backgrounds,  an
upper limit is derived. The 
limit determination is performed,  in the \textsc{RooStats} framework~\cite{Moneta:2010pm},  using the asymptotic CL$_{s}$
method~\cite{cowan}. This is  an approximate method, equivalent to the
true CL$_{s}$ method performed with simulated pseudo-experiments, when
the data samples are not too small.

\begin{figure}[!htbp]
\begin{center}
\begin{overpic}[width=
0.69\textwidth]{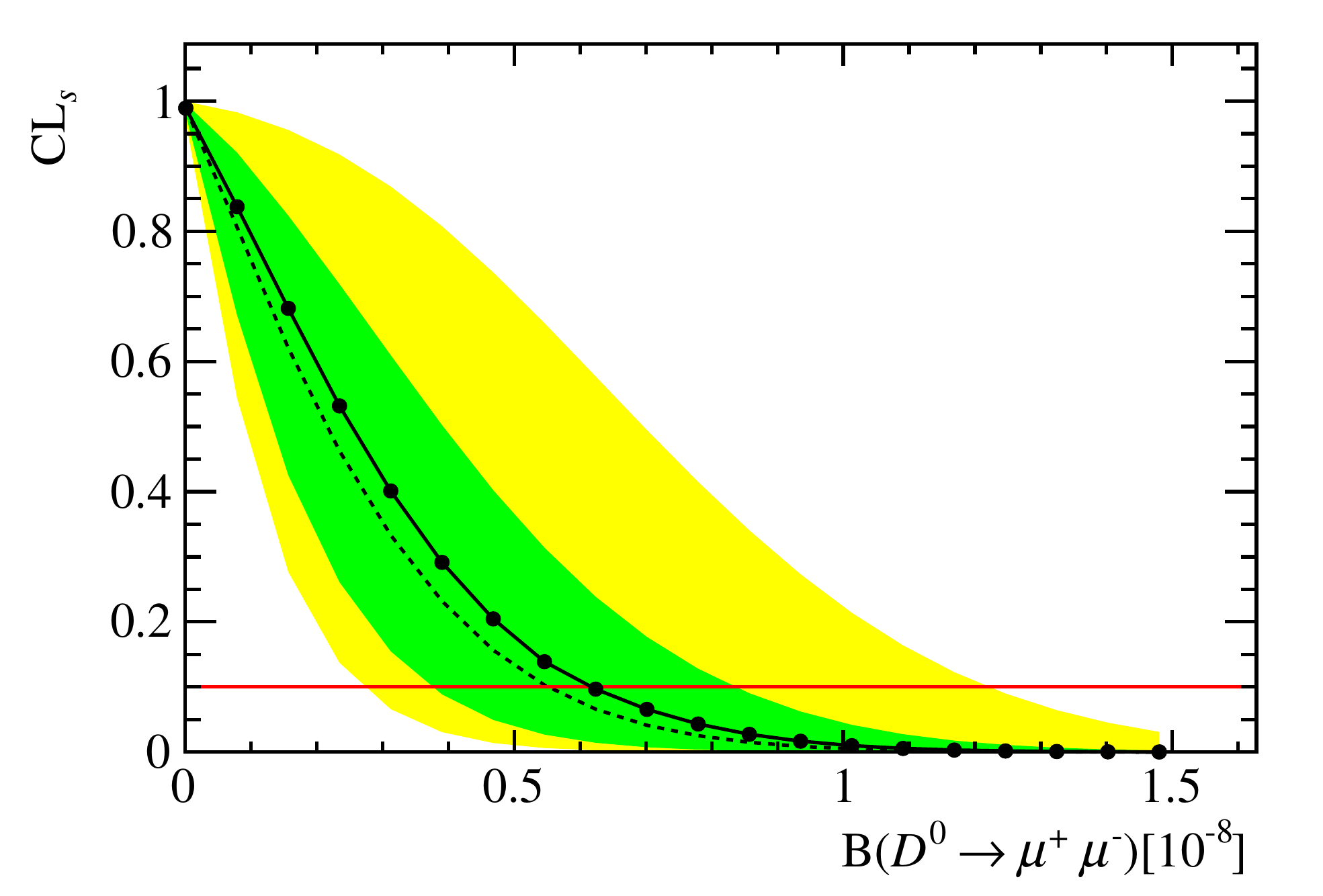}
\put(80,50){LHCb}
\end{overpic}  
\end{center}
\caption{\small   CL$_{s}$ (solid line) as a function of the assumed \dmumu branching fraction
and  median (dashed line), 1$\sigma$
and 2$\sigma$ bands of  the expected CL$_{s}$, in the background-only
hypothesis, obtained with the asymptotic CL$_{s}$
method. The horizontal line corresponding to CL$_{s}$=0.05 is also drawn.
}\label{fig:limit}
\end{figure}

Figure~\ref{fig:limit} shows the expected and observed CL$_{s}$ as a function of
the assumed \mbox{\dmumu} branching fraction.
The expected upper limit is
\mbox{$ 5.5 \: (6.7)^{+3.1}_{-2.0} \times 10^{-9}$} at \mbox{90\% }  \mbox{(95\%) CL}, while the  observed limit is
\mbox{$ 6.2 \: (7.6)   \times 10^{-9}$} at \mbox{90\% }  \mbox{(95\%) CL}. The p-value 
for the  background-only hypothesis is 0.4.

The robustness of the fit procedure is tested with simulated
pseudo-experiments  using the same starting values for the fit
parameters used in  the  data fit except for the combinatorial background PDF,
for which the fitted parameters from  data  are used. 
Simulated pseudo-experiments  are performed corresponding to \dmumu
branching fraction  values of    0, $10^{-8}$ and $5\times 10^{-8}$. In
all cases the results reproduce the input values within the estimated uncertainties. 

Several systematic checks are performed varying the selection requirements,
including the muon identification criteria,  varying the parametrization of the fit
components and the fit range and removing the multivariate selection.  The measured \brmumu
does not change significantly with these variations.

To test the dependence of the result on the knowledge of  the double
misidentification  probability, the uncertainty is doubled in the fit
input; \brmumu  is consistent with the baseline result.

In addition, the robustness of the result is  checked by artificially
increasing the value  of  the kaon to muon misidentification as
determined from data in Section~\ref{sec:misid} up to 200\% of its measured
value,  and the fitted branching fraction still remains   consistent with no significant
excess of signal  with respect to the  background expectations.

\section{Summary}
\label{sec:results}

A search for the rare decay \dmumu  is performed using a data sample, corresponding to an
integrated luminosity of  0.9~\invfb, of 
  $pp$ collisions collected at
a centre-of-mass energy of  7 TeV  by the LHCb experiment. The observed number of events is consistent with the
background expectations and  corresponds to an  upper limit of
\begin{equation*}
{\cal B}(\dmumu)<  6.2 \: (7.6)  \times 10^{-9} {\rm \: at  \: 90\% \: (95\%)\: CL.} 
\end{equation*}
 This result represents an improvement of more than a factor twenty with respect
 to previous measurements but remains several orders of magnitude larger than the 
SM prediction.

 \section*{Acknowledgements}

\noindent We express our gratitude to our colleagues in the CERN
accelerator departments for the excellent performance of the LHC. We
thank the technical and administrative staff at the LHCb
institutes. We acknowledge support from CERN and from the national
agencies: CAPES, CNPq, FAPERJ and FINEP (Brazil); NSFC (China);
CNRS/IN2P3 and Region Auvergne (France); BMBF, DFG, HGF and MPG
(Germany); SFI (Ireland); INFN (Italy); FOM and NWO (The Netherlands);
SCSR (Poland); ANCS/IFA (Romania); MinES, Rosatom, RFBR and NRC
``Kurchatov Institute'' (Russia); MinECo, XuntaGal and GENCAT (Spain);
SNSF and SER (Switzerland); NAS Ukraine (Ukraine); STFC (United
Kingdom); NSF (USA). We also acknowledge the support received from the
ERC under FP7. The Tier1 computing centres are supported by IN2P3
(France), KIT and BMBF (Germany), INFN (Italy), NWO and SURF (The
Netherlands), PIC (Spain), GridPP (United Kingdom). We are thankful
for the computing resources put at our disposal by Yandex LLC
(Russia), as well as to the communities behind the multiple open
source software packages that we depend on.

\bibliographystyle{LHCb}
\bibliography{main}

\end{document}